\documentclass[twocolumn,superscriptaddress,showpacs,prl]{revtex4}
\usepackage{graphicx}
\usepackage{dcolumn}
\usepackage{natbib}
\usepackage{color}
\usepackage{float}

\newcommand{\kpp}{K^0_L\rightarrow\pi^0\pi^0}
\newcommand{\kppp}{K^0_L\rightarrow 3\pi^0}

\newcommand{\spmm}{\Sigma^+ \rightarrow p \mu^+ \mu^-}
\newcommand{\kppx}{K^0_L \rightarrow \pi^0 \pi^0 X}
\newcommand{\xgg}{X \rightarrow \gamma \gamma}

\newcommand*{\TAIWAN}{%
Department of Physics, National Taiwan University, Taipei, Taiwan 10617 Republic of China}
\newcommand*{\PUSAN}{%
Department of Physics, Pusan National University, Busan, 609-735 Republic of Korea}
\newcommand*{\SAGA}{%
Department of Physics, Saga University, Saga, 840-8502 Japan}
\newcommand*{\DUBNA}{%
Laboratory of Nuclear Problems, Joint Institute for Nuclear Research, 
Dubna, Moscow Region, 141980 Russia}
\newcommand*{\SOKENDAI}{%
Department of Particle and Nuclear Research, 
The Graduate University for Advanced Science (SOKENDAI), Tsukuba, Ibaraki, 305-0801 Japan}
\newcommand*{\KEK}{%
Institute of Particle and Nuclear Studies, 
High Energy Accelerator Research Organization (KEK), Tsukuba, Ibaraki, 305-0801 Japan}
\newcommand*{\OSAKA}{%
Department of Physics, Osaka University, Toyonaka, Osaka, 560-0043 Japan }
\newcommand*{\YAMAGATA}{%
Department of Physics, Yamagata University, Yamagata, 990-8560 Japan}
\newcommand*{\CHICAGO}{%
Enrico Fermi Institute, University of Chicago, Chicago, Illinois 60637, USA }
\newcommand*{\NDA}{%
Department of Applied Physics, National Defense Academy, Yokosuka, Kanagawa, 239-8686 Japan}
\newcommand*{\RCNP}{%
Research Center of Nuclear Physics, Osaka University, Ibaraki, Osaka, 567-0047 Japan}
\newcommand*{\KYOTO}{%
Department of Physics, Kyoto University, Kyoto, 606-8502 Japan}
\newcommand*{\IHEP}{%
Institute for High Energy Physics, Protvino, Moscow region, 142281 Russia}
\newcommand*{\GOMEL}{%
Scarina Gomel' State University, Gomel', BY-246699, Belarus}
\newcommand*{\ARIZONA}{%
Department of Physics and Astronomy, Arizona State University, Tempe, Arizona, USA}
\newcommand*{\RIKEN}{%
RIKEN SPring-8 Center, Sayo, Hyogo, 679-5148 Japan}
\newcommand*{\NY}{%
University  of Rochester, Rochester, NY 14627}
\newcommand*{\CERN}{%
CERN, CH-1211 Gen\`{e}ve 23, Switzerland}
\begin{document}
\date{24 October 2008}
\title{
Search for a light pseudoscalar particle in the decay $K^0_L \rightarrow \pi^0 \pi^0 X$ }

\author{Y.~C.~Tung}\affiliation{\TAIWAN}
\author{Y.~B.~Hsiung}\affiliation{\TAIWAN}
\author{M.~L.~Wu}\affiliation{\TAIWAN}
\author{K.~F.~Chen}\affiliation{\TAIWAN}
\author{J.~K.~Ahn}\affiliation{\PUSAN} 
\author{Y.~Akune}\affiliation{\SAGA} 
\author{V.~Baranov}\affiliation{\DUBNA}
\author{J.~Comfort}\affiliation{\ARIZONA} 
\author{M.~Doroshenko}\altaffiliation{Present address: \DUBNA}\affiliation{\SOKENDAI} 
\author{Y.~Fujioka}\affiliation{\SAGA} 
\author{T.~Inagaki}\affiliation{\SOKENDAI}\affiliation{\KEK} 
\author{S.~Ishibashi}\affiliation{\SAGA}
\author{N.~Ishihara}\affiliation{\KEK}
\author{H.~Ishii}\affiliation{\OSAKA} 
\author{E.~Iwai}\affiliation{\OSAKA}
\author{T.~Iwata}\affiliation{\YAMAGATA} 
\author{I.~Kato}\affiliation{\YAMAGATA} 
\author{S.~Kobayashi}\affiliation{\SAGA}
\author{T.~K.~Komatsubara}\affiliation{\KEK} 
\author{A.~S.~Kurilin}\affiliation{\DUBNA} 
\author{E.~Kuzmin}\affiliation{\DUBNA}
\author{A.~Lednev}\affiliation{\IHEP}\affiliation{\CHICAGO} 
\author{H.~S.~Lee}\affiliation{\PUSAN} 
\author{S.~Y.~Lee}\affiliation{\PUSAN} 
\author{G.~Y.~Lim}\affiliation{\KEK}
\author{J.~Ma}\affiliation{\CHICAGO}
\author{T.~Matsumura}\affiliation{\NDA}
\author{A.~Moisseenko}\affiliation{\DUBNA}
\author{H.~Morii}\affiliation{\KYOTO} 
\author{T.~Morimoto}\affiliation{\KEK}
\author{T.~Nakano}\affiliation{\RCNP} 
\author{H.~Nanjo}\affiliation{\KYOTO}
\author{J.~Nix}\affiliation{\CHICAGO}
\author{T.~Nomura}\altaffiliation{Present address: \KEK}\affiliation{\KYOTO}
\author{M.~Nomachi}\affiliation{\OSAKA}
\author{R.~Ogata}\affiliation{\SAGA}
\author{H.~Okuno}\affiliation{\KEK}
\author{K.~Omata}\affiliation{\KEK}
\author{G.~N.~Perdue}\altaffiliation{Present address: \NY}\affiliation{\CHICAGO} 
\author{S.~Podolsky}\altaffiliation{Present address: \GOMEL}\affiliation{\DUBNA} 
\author{K.~Sakashita}\altaffiliation{Present address: \KEK}\affiliation{\OSAKA} 
\author{T.~Sasaki}\affiliation{\YAMAGATA} 
\author{N.~Sasao}\affiliation{\KYOTO}
\author{H.~Sato}\affiliation{\YAMAGATA}
\author{T.~Sato}\affiliation{\KEK}
\author{M.~Sekimoto}\affiliation{\KEK}
\author{T.~Shinkawa}\affiliation{\NDA}
\author{Y.~Sugaya}\affiliation{\OSAKA}
\author{A.~Sugiyama}\affiliation{\SAGA}
\author{T.~Sumida}\altaffiliation{Present address: \CERN}\affiliation{\KYOTO}
\author{S.~Suzuki}\affiliation{\SAGA}
\author{Y.~Tajima}\affiliation{\YAMAGATA}
\author{S.~Takita}\affiliation{\YAMAGATA} 
\author{Z.~Tsamalaidze}\affiliation{\DUBNA}
\author{T.~Tsukamoto}\altaffiliation{Deceased.}\affiliation{\SAGA} 
\author{Y.~Wah}\affiliation{\CHICAGO}
\author{H.~Watanabe}\altaffiliation{Present address: \KEK}\affiliation{\CHICAGO}
\author{M.~Yamaga}\altaffiliation{Present address: \RIKEN}\affiliation{\KEK}
\author{T.~Yamanaka}\affiliation{\OSAKA}
\author{H.~Y.~Yoshida}\affiliation{\YAMAGATA}
\author{Y.~Yoshimura}\affiliation{\KEK}
\author{Y.~Zheng}\affiliation{\CHICAGO}

\collaboration{E391a Collaboration}\noaffiliation

\begin{abstract}
We performed a search for a light pseudoscalar particle $X$
in the decay $\kppx$, $\xgg$ with the E391a detector at KEK.
Such a particle with a mass of 214.3 MeV/$c^2$ was suggested by the HyperCP experiment.
We found no evidence for $X$ and 
set an upper limit on the product branching ratio for $\kppx$, $\xgg$ of $2.4 \times 10^{-7}$
at the $90\%$ confidence level. 
Upper limits  on the branching ratios in the mass
region of $X$ from $194.3$ to $219.3$ MeV/$c^2$ are also presented.

\end{abstract}

%
%
%
\maketitle
%

We report the results of a search for the decay $\kppx$, $\xgg$, 
where $X$ is a pseudoscalar particle of mass in the region $194.3-219.3$ MeV/$c^2$.
No experimental study for $\xgg$ has been published in this mass range.
This study was motivated by the three events of the decay $\spmm$ with the dimuon
invariant mass around 214.3 MeV/$c^2$ reported by the HyperCP collaboration in 2005~\cite{hypercp}.

A sgoldstino interpretation~\cite{ds1,ds2} for the HyperCP observation
showed an upper estimate of the branching ratio, 
but this calculation strongly depends on a complex coupling constant
and only the absolute value can be extracted from HyperCP data.
The model also showed that the $X$ particle's branching ratio (BR)
should be saturated by the channels 
$X \rightarrow \gamma \gamma $ and $X \rightarrow \mu^+ \mu^-$, 
with BR$\left(\xgg\right)$/BR$\left(X \rightarrow \mu^+ \mu^-\right)$ $\sim 10^{4}$.
Another model~\cite{he} suggested a light pseudoscalar Higgs boson interpretation of $X$,
but there was no prediction for $\kppx,$ $\xgg$ decay BR.

We searched for the $\kppx$, $\xgg$ decay at the KEK E391a experiment~\cite{e391run2}.
Neutral kaons were produced by 12~GeV protons
incident on a 0.8-cm-diameter and 6-cm-long platinum target.
The proton intensity was typically $2 \times 10^{12}$
per spill coming every 4 sec.
The neutral beam~\cite{beamline}, 
with a solid angle of 12.6~$\mu$str,
was defined by a series of six sets of collimators and a pair of sweeping magnets 
aligned at a production angle of 4~degrees. 
A 7-cm-thick lead block and a 30-cm-thick beryllium block were placed
between the first and second collimators
to reduce beam photons and neutrons.
The beam size at $11.8$~m downstream of the target,
measured with the E391a detector,
was $3.7$ cm (FWHM) including the effects of detector resolution.
The beam line was kept in vacuum 
at 1 Pa after 5~m downstream of the target
and $1 \times 10^{-5}$ Pa inside the fiducial decay region.
The $K^0_L$ momentum peaked around 2~GeV/$c$ at the entrance of the detector,
11.8~m downstream of the target.

\begin{figure}[b]
   \includegraphics[angle=-90, width=8.6cm]{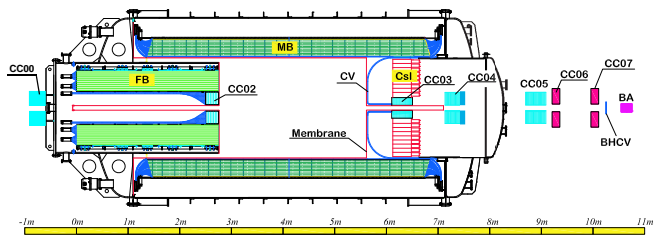}
   \caption{\label{fig:det1}%
	(color online) 
	Schematic cross-sectional view of the E391a detector.
	``0m'' in the scale corresponds to 
	the entrance of the front barrel (FB) detector.
	$K^0_L$'s entered from the left side.
	}
\end{figure}

Figure~\ref{fig:det1} shows a cross-sectional view of the E391a detector
and defines the origin of our coordinate system.
 The detector components were cylindrically assembled
 along the beam axis. Most were installed inside the vacuum tank 
 to minimize interactions of the particles before detection. 
The electromagnetic calorimeter, labeled ``CsI'',
 measured the energy and position of the photons from $\pi^0$ and $X$ decays.
It consisted
 of 496 blocks of \(7 \times 7 \times 30~\mbox{cm}^3\)
 undoped CsI crystal and 80 specially shaped CsI blocks 
 used in the peripheral region, 
 covering a circular area with a 95 cm radius.
To allow beam particles to pass through, 
the calorimeter had a $12 \times 12$~cm$^2$ 
beam hole at the center.
The main barrel (MB) and 
 front barrel (FB) counters consisted of 
 alternating layers of lead and scintillator sheets with 
total thicknesses of 13.5~$X_0$ and 17.5~$X_0$, respectively,
and surrounded the neutral beam.
Both upstream and downstream ends of MB and the upsteam end of FB 
had Hamamatsu R329-EGP photomultiplier tubes~\cite{pmt} for scintillation light detection.
To identify charged particles entering the calorimeter, 
a scintillation counter (CV) hermetically covered the front of the calorimeter. 
It consisted of a plastic scintillator hodoscope placed 50~cm upstream of the calorimeter 
with a $12\times12$~cm$^2$ beam hole at the center,
and four 6~mm thick scintillator plates 
that connected the beam holes in the hodoscope and the calorimeter. 
 Multiple collar-shaped photon counters
 (CC00, CC02--07) were placed 
 along the beam axis
to detect particles escaping 
 in the beam direction.
The CC02 was a shashlik type lead-scintillator sandwich counter 
with optical fibers running perpendicularly 
to the lead and scintillator plates
through aligned holes,
and was located at the upstream end of the $K^0_L$ decay region.
The CC03 filled the volume between the beam hole 
and the innermost layers of the CsI blocks in the calorimeter. 
 The vacuum region was separated by a thin multi-layer film (``membrane'') 
 between the beam and detector regions.
This kept the decay region at  $1 \times 10^{-5}$~Pa despite some outgassing from the detector.
Further descriptions of the E391a detector are given in \cite{e391run2,detector}.

 In this analysis, 
 we used data taken in the run period from February to April 2005, or Run-II. 
Data were taken with a hardware trigger requiring two or 
more shower clusters in the calorimeter with cluster energy $\ge 60$~MeV. 
We also required no activity in the CV and in some other photon counters. 
Because both the decays $\kppx$, $\xgg$ and $\kppp$ have a signature of six photons in the final state,
separation of these decays was crucial in this study.

In the analysis,
the $\kppp$ and $\kppx$ decays were simulated using GEANT3 Monte Carlo (MC) 
simulations~\cite{geant}
and were overlaid with accidental events 
taken from the target-monitor accidental trigger.
In the $\kppx$ decay, $X$ was assumed to decay immediately to two photons.
To reconstruct $\kppx$, 
we selected events with six photon-like clusters on the CsI calorimeter 
without any in-time hits on the other detectors.
All the clusters were required to be within the fiducial region,
which was outside the 25 cm by 25 cm square around the beamhole 
and inside a 88 cm circle from the center of the beamline.
An additional selection criterion on the transverse momentum of $K_L^0$ ($P_T$ $<$ 0.025 GeV/$c$) 
was required to suppress missing photon events.
Afterwards, event reconstruction proceeded by solving for the decay vertex, 
assuming the $\pi^0$ mass and 
constraining the vertex to lie along the beam axis.
The fiducial decay Z-vertex ($Z_{vtx}$) region was defined to be between 250 and 550 cm.
There were 45 possible combinations to select two photon pairs from 6 photons to form two $\pi^0$'s.
The most likely pairing was chosen by a minimum $\chi^2_z$,
which was calculated based on the difference between reconstructed $Z_{vtx}$'s of the $\pi^0$'s.
In the $\kppx$ event,
the $X$ mass was the invariant mass of the third photon pair $\left(M_{56}\right)$
and was reconstructed from the energy and hit positions of the remaining two photons 
and the $Z_{vtx}$ determined from two $\pi^0$'s.
\begin{figure}[htb] 
   \includegraphics[width=8.6cm]{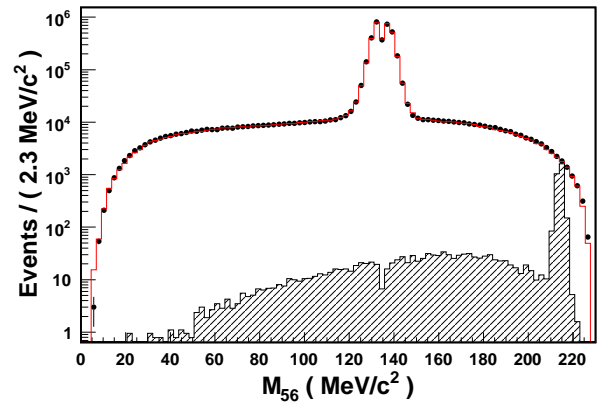}
   \caption{
   \label{fig:rec}%
  (color online)
	The $M_{56}$ distribution of the events with six photons in the calorimeter. 
  The points with error bars show the data, 
  and the red (open) histogram shows the $\kppp$ MC results normalized to 
  the number of data events.
  The shaded histogram represents the $\kppx$ MC results ($m_X=214.3$ MeV/$c^2$) and was normalized assuming 
  that $BR\left(\kppx,~X \rightarrow \gamma \gamma\right)$ is $1.2\times10^{-4}$.
	}
\end{figure}
The event reconstruction was further optimized by a constrained fit~\cite{fit} 
requiring:
(a) six photons to have the $K_L^0$ mass,
(b) two photon pairs to have the $\pi^0$ mass, and
(c) the $K_L^0$ momentum vector from the target to point to the center of energy of
   the photons on the calorimeter.
The $\chi^2$ of the constrained fit ($\chi^2_{fit}$) was calculated for  all 45 combinations,
and the one with the minimum $\chi^2_{fit}$ was chosen to be the correct pairing.
For successful reconstructions, 
$M_{56}$ equals the $\pi^0$ mass for $\kppp$ decays 
and the $X$ mass ($m_X$) for $\kppx$ decays. 
Figure~\ref{fig:rec} shows $M_{56}$ of data, 
the $\kppp$ MC results and the $\kppx$ MC results with a $214.3$ MeV/$c^2$ $X$ mass.
The signal region is defined to be 211.3 MeV/$c^2$ $<$ $M_{56}$ $<$ 217.3 MeV/$c^2$. 
Since the $\kppp$ decay has three $\pi^0$'s in the final state, 
$M_{56}$ could be the invariant mass of any one of the three $\pi^0$'s.
Since only the reconstructed masses of the first and second photon pairs 
are constrained,
the pairing with the minimum $\chi^2_{fit}$
left the worst reconstructed mass of $\pi^0$ in the third photon pair.
This explains 
the dip in the $\pi^0$ peak.

\begin{figure}[htb]
   \includegraphics[width=8.6cm]{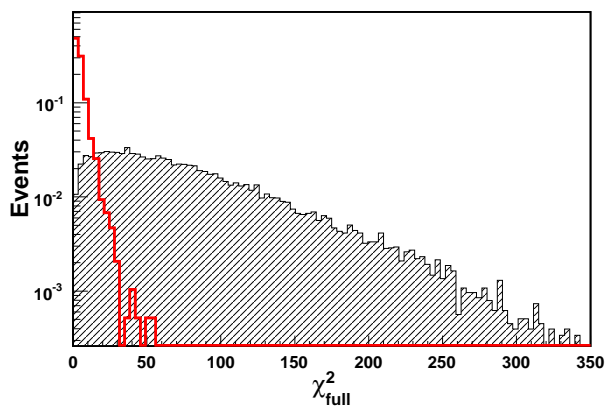}
   \caption{\label{fig:chi2}%
   (color online)
   The $\chi^2_{full}$ distribution of events
   just before the $\chi^2_{full}$ cut, explained in the text,
   in the signal region
   of $m_X=214.3$ MeV/$c^2$.
   The red (open) histogram shows the $\kppp$ MC results 
   and shaded histogram shows the $\kppx$ MC results ($m_X=214.3$ MeV/$c^2$ ). 
   The cut point was set at 50.
}
\end{figure}

The $\kppp$ mode, with the $19.56\%$ branching ratio~\cite{PDG},
was our dominant background source. 
Our $\kppp$ MC results show
the tail of the $\pi^0$ mass peak extended into the $X$ mass region.
This background was caused by wrong photon pairing combinations,
and was suppressed by requiring 
$\chi^2_z$ $<$ 4 and $\chi^2_{fit}$ $<$ 6.
The background was further suppressed by rejecting events
consistent with $\kppp$ decays.
We applied another constrained fit, called the ``full constrained fit'',
by requiring the three photon pairs to have $m_{\pi^0}$,
and reconstructing each event as $\kppp$.
As shown in Fig.~\ref{fig:chi2}, $\kppp$ events in the signal region
have smaller $\chi^2$ ($\chi^2_{full}$) than $\kppx$.
By discarding events with $\chi^2_{full}$ $<$ 50, 
the wrong pairing background events in the region $M_{56}$ $>$ 165 MeV/$c^2$
were suppressed by a factor $6.6 \times 10^{2}$.
The acceptance of $\kppx$ ($m_X=214.3$ MeV/$c^2$ ) decays 
by the $\chi^2_{full}$ cut was estimated to be $61.0\%$
based on the simulations.

\begin{figure}[hb]
   \includegraphics[width=8.6cm]{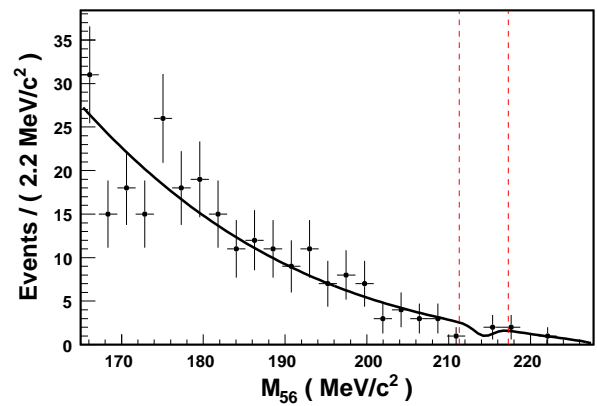}
   \caption{\label{fig:final}%
	The modeling result of the 214.3 MeV/$c^2$ $X$ mass. 
	Error bars show the data, and the solid line
	which is the combination of background and signal PDF, 
	is the modeling result. 
	The region between two dashed lines is the signal mass region.
	}
\end{figure}

With all the selection cuts applied on the data, 
only 2 events survived in 
the 6-MeV/$c^2$ wide $X$ mass region centered on 214.3 MeV/$c^2$
and 250 events in sideband regions
(165~MeV/$c^2$ $<$ $M_{56}$ $<$ 211.3~MeV/$c^2$ and $M_{56}$ $>$ 217.3~MeV/$c^2$).
The surviving events 
had a distribution consistent with $\kppp$ decays.
We adopted an unbinned extended-likelihood method to extract the number of signal events.
Since the $X$ mass was close to the kinematical boundary at $227.7$ MeV/$c^2$, 
the background shape was modeled by the events in data sideband with the ARGUS function \cite{argus},
which is defined as:
\begin{eqnarray}
	\nonumber
   f\left(x\right)_{ARGUS}= x\sqrt{1-\left({\frac{x}{M_{ep}}}\right)^2}\exp\left[{\alpha-\alpha\left({\frac{x}{M_{ep}}}\right)^2}\right],
\end{eqnarray}
where $M_{ep}$ is the end point of the ARGUS function, $x$ is $M_{56}$, 
and $\alpha$ determines the curvature of ARGUS function.
For the signal MC results, 
the reconstructed mass peak is asymmetric because of the kinematical limit of the $X$ mass.
In this case, 
a double Gaussian function with different mean values of the two composition Gaussian functions
was chosen as the probability density function (PDF) to model the signal shape from the signal MC results.
The modeling result of the 214.3 MeV/$c^2$ $X$ mass is shown in Fig.~\ref{fig:final}.
The number of signal events obtained from the extended-likelihood method,
$N_s$, was $-1.4^{+1.7}_{-0.9}$.
A dip around 214.3 MeV/$c^2$ shows that a negative number of signal events was yielded.
In this paper, we searched for $X$ in the mass region $>$ 190 MeV/$c^2$ and no evidence for $X$ was found.
We have already published an upper limit
on the branching ratio of $X$ decaying into invisible particles,
in the low mass region ($m_X$ $<$ 100 MeV/$c^2$)~\cite{jnix}.

%

%
\begin{table}[ht]
   \caption{\label{table:BG}%
	A summary of the systematic uncertainties. 
	The uncertainty in the background(bkgd) shape is shown in the uncertainty of $N_s$ 
	and the others are shown in the percentage uncertainty on the branching ratio.	
	}
   \begin{ruledtabular}
   \begin{tabular}{lcccccc}
$m_X$ (MeV/$c^2$) & $194.3$ & $199.3$ & $204.3$ & $209.3$ & $214.3$ & $219.3$~\\
\hline
$(a)~Sig.~shape$~($\%$)					&	$6.37$	&	$3.39$	&	$8.62$	&	$9.54$	&	$4.07$	&	$0.23$~\\
$(b)~Sig.~resolution$~($\%$) 		&	$1.36$	&	$0.29$	&	$2.42$	&	$2.64$	&	$1.79$	&	$0.56$~\\
$(c)~Sig.~efficiency$~($\%$) 		&	$0.93$	&	$0.93$	&	$0.93$	&	$0.93$	&	$0.93$	&	$0.93$~\\
$(d)~K_L^0~flux$~($\%$) 				&	$2.88$	&	$2.88$	&	$2.88$	&	$2.88$	&	$2.88$	&	$2.88$~\\
$(e)~Bkgd~shape$~($\Delta N_s$)	&	$1.12$ 	&	$1.42$	&	$2.35$	&	$1.64$	&	$0.80$	&	$0.27$~\\
   \end{tabular}
   \end{ruledtabular}
\end{table}
\begin{table*}[htpb]
\caption{\label{table:Results}%
A summary of the signal yields ($N_s$), the acceptances of the signal decay (Accept.), single event sensitivities (S.E.S.), 
central value of the branching ratios (B.R.), and the upper limits at the $90\%$ confidence level(U.L.).
}
\begin{center}
\begin{ruledtabular}
\begin{tabular}{cccccc}
$m_X$ (MeV/$c^2$)~ & $N_s$ & Accept. ($10^{-4}$)  & S.E.S. ($10^{-8}$)  & B.R. ($10^{-8}$) & U.L. ($10^{-7}$) \\ 
\hline
$194.3$	&  $6.4^{+6.0}_{-5.2}$~  	& $1.09\pm0.01$ & $7.0\pm1.3(syst.)$  	&  $44.5^{+41.6}_{-36.4}(stat.)		\pm8.4(syst.)$~ & 10.7~ \\
$199.3$ &  $3.5^{+4.8}_{-4.0}$~  	& $1.09\pm0.01$ & $7.0\pm2.9(syst.)$  	&  $24.1^{+33.2}_{-27.8}(stat.)		\pm10.0(syst.)$~ & 7.9~ \\
$204.3$ &  $-0.7^{+3.4}_{-2.7}$~  &	$1.09\pm0.01$ & $7.0\pm23.0(syst.)$  	&  $-4.9^{+23.9}_{-18.8}(stat.)		\pm16.3(syst.)$~ & 5.0~ \\
$209.3$ &  $-1.5^{+2.5}_{-1.8}$~ 	& $1.08\pm0.01$ & $7.0\pm7.6(syst.)$ 		&  $-10.7^{+17.2}_{-12.3}(stat.)	\pm11.6(syst.)$~ & 3.5~ \\
$214.3$ &  $-1.4^{+1.7}_{-0.9}$~ 	&	$1.08\pm0.01$ & $7.0\pm4.0(syst.)$  	&  $-10.0^{+11.8}_{-6.6}(stat.)		\pm5.6(syst.)$~ & 2.4~ \\
$219.3$ &  $-0.3^{+1.7}_{-1.0}$~ 	&	$1.09\pm0.01$ & $7.0\pm5.6(syst.)$  	&	 $-2.4^{+12.0}_{-6.5}(stat.)		\pm1.9(syst.)$~ & 2.6~ \\
\end{tabular}
\end{ruledtabular}
\end{center}
\end{table*} 
Table~\ref{table:BG} summarizes systematic uncertainties due to
$\left(a\right)$ the signal shape modeling,
$\left(b\right)$ the signal resolution difference between data and the MC results,
$\left(c\right)$ the signal efficiency,
$\left(d\right)$ the $K_L^0$ flux estimation, and
$\left(e\right)$ the background shape modeling.
The dominant uncertainty was source $\left(e\right)$.
The uncertainty $\left(a\right)$ was evaluated by comparing
the value of $N_s$
while changing the signal PDF width by 1$\sigma$ fitted error in the signal extraction.
The width of the 214.3-MeV/$c^2$ $X$ mass distribution modeled
by a single Gaussian function was 1.2 MeV/$c^2$. 
The uncertainties $\left(b\right)$ and $\left(e\right)$ were evaluated in similar way by 
changing the width of the signal PDF and the curvature of the background PDF.
Since the signal peak was not found in data, 
the uncertainty $\left(b\right)$ was evaluated indirectly by 
assuming that the difference in signal resolution is 
the same as the $\pi^0$ resolution in reconstructed $\kppp$ between data and the MC results.
The $\pi^0$ mass peak, which had a hollow dip in the center of the peak, 
was modeled by the double Gaussian function 
composed of the two Gaussian functions with opposite norm.
The one with positive norm described the sides of the peak 
and the one with negative norm took out the hollow area in the center.
The percentage difference of the resolution between data and the MC results
was estimated to be $1.28\%$.
During signal extraction the background PDF was fixed.
The uncertainty $\left(e\right)$ was evaluated by comparing the change in $N_s$ 
while floating the curvature of the ARGUS function 
in the signal extraction.
The uncertainty $\left(c\right)$ was simply determined by the statistics of the signal MC results.
The number of accumulated $K_L^0$ was estimated by the $\kppp$ mode 
and was cross-checked by the $\kpp$ mode~\cite{PDG}.
Acceptance discrepancies in selection cuts 
between data and MC results create an uncertainty in the $K_L^0$ flux estimation.  
That uncertainty is listed in row~(d).
The $K_L^0$ flux at 10 m from the target was determined to be
$\left(1.32\pm0.04\right)\times10^{11}$
based on the number of decays downstream of that point.

The results of different $m_X$ are summarized in Table~\ref{table:Results}.
$N_s$ was the number of signal events yielded, and the quoted error was the fitting error.
The signal acceptance was calculated using MC code.
From the acceptance and the $K_L^0$ flux, 
the single event sensitivity for $\kppx$, $\xgg$ was defined as:
\begin{eqnarray}
	\nonumber
    S.E.S.(\kppx) = \frac{1}{\mbox{Acceptance}\cdot N(K^0_L \mbox{decays})} \;.
\end{eqnarray}
The error quoted in S.E.S. was evaluated by summing all the systematic uncertainties quadratically.
The negative central value of the branching ratio was due to the negative yield of $N_s$ in the signal extraction.
The upper limit was calculated by 
integrating up to $90\%$ of the area under the likelihood function,
for $N_s$ $>$ 0.
The systematic uncertainties were incorporated by 
convolving the likelihood function with a Gaussian function, 
and the statistical uncertainty was incorporated while integrating the likelihood function.
\begin{figure}[ht]
   \includegraphics[width=8.6cm]{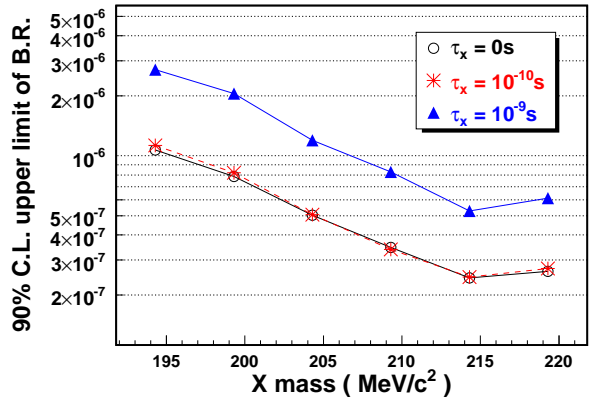}
   \caption{\label{fig:upper}%
	(color online) 
	The $90\%$ confidence level upper limits of the BR($\kppx$, $\xgg$) 
	for different $X$ lifetimes as a function of the $X$ mass.
	}
\end{figure}
The upper limit on the branching ratio for $\kppx$, $\xgg$ decay 
depends on the lifetime ($\tau_X$) and the mass of $X$ as shown in Fig.~\ref{fig:upper}.
The upper limits were independent of the $X$ lifetime if it was shorter than $10^{-10}$s.
The upper limit on the branching ratio for $\kppx$, $\xgg$ ($m_X=214.3$ MeV/$c^2$) decay
was set to be $2.4\times10^{-7}$ in the lifetime region of $\tau_X < 10^{-10}s$.
The upper limits weaken by a factor of 2--3 if the $X$ lifetime was $10^{-9}$s.
The upper bound of the $X$ lifetime was estimated to be $2.5\times10^{-11}$s by~\cite{ds1}.
%

\begin{acknowledgements}
We are grateful to the operating crew of the KEK 12-GeV proton synchrotron 
for their successful beam operation during the experiment. 
This work has been partly supported by a Grant-in-Aid from the MEXT and JSPS in Japan, 
a grant from National Science Council in Taiwan, from the U.S. Department of Energy
and from Korea Research Foundation.
\end{acknowledgements}

\bibliographystyle{plain}

\end{document}